\documentclass[twocolumn,aps,epsfig]{revtex4}
\usepackage{graphicx}
\usepackage{epstopdf}
\usepackage{latexsym}
\usepackage{amssymb}
\usepackage[center]{subfigure}
\begin{document}
 \newcommand{\bq}{\begin{equation}}
 \newcommand{\eq}{\end{equation}}
 \newcommand{\bqn}{\begin{eqnarray}}
 \newcommand{\eqn}{\end{eqnarray}}
 \newcommand{\nb}{\nonumber}
 \newcommand{\lb}{\label}
\title{A Five Dimensional Space Without Local Lorentz Invariance}

\author{Ahmad Borzou}
\email{Ahmad_Borzou@baylor.edu}

\affiliation{Physics Department, Baylor
University, Waco, TX 76798-7316, USA}

\date{\today}

\begin{abstract}
 A five dimensional space without invariance under local Lorentz transformations is studied, and the transformations under which the theory is invariant are introduced. We show that the Lorentz force is included in the ensuing equations of motion. The four dimensional Einstein and Maxwell equations emerge from the field equations while the homogeneous Maxwell equations naturally appear in the work. A corresponding quantum theory is introduced. The spectrum of a free particle and the hydrogen atom are recovered. Quantum structure of Schwarzschild spaces are also studied.

\end{abstract}

\maketitle

\section{Introduction}
After Einstein published the general theory of relativity in 1915, numerous attempts were made to generalize the theory in such a way as to encompass both gravitation and electromagnetism in a unique geometrical structure. Einstein and his collaborators tried to unify gravity and electromagnetism within a generalization of vacuum general relativity which allows for non-symmetric fields \cite{1, 2, 3, 4, 5, 6, 7, 8}. Weyl introduced a new geometry in which a vector field and metric tensor characterize space-time \cite{9}. Eddington developed a geometry within which connection rather than metric is fundamental field \cite{eddington}. Perhaps, the work of Kaluza and Klein is more famous among others. Their theory was five dimensional general relativity \cite{kaluza, klein}. One can get good reviews on Kaluza-Klein theories in \cite{Bailint, wesson}. Many other works have been done. In \cite{hubert} Goenner has reviewed most of them. There is as yet no clear and consistent picture of the subject. On the other hand, most of physicists including Einstein believe that all the physical laws are from a single origin. Therefore, the necessity for finding at least a classical unified field theory is obvious. Since any classical theory is a limited version of a more basic quantum one, finding this classical theory would be probably a good spot to develop a more fundamental corresponding quantum theory. The theory we now know as quantum mechanics faces what Von Neumann \cite{neumann} calls ``chief weakness" of the theory. This theory, even in its relativistic version, distinguishes time from the three spatial coordinates which are presented as operators, and, therefore, has a non-relativistic character.

In the Kaluza-Klein theories people usually impose a constraint on the Faraday tensor to get the two homogeneous Maxwell equations. This can not be considered as a kind of unification. One of the motivations of the following work is to derive not only the non-homogeneous but also the homogeneous Maxwell equations from a unified theory of gravity and electromagnetism. By finding a classical unified field theory, we also hope to be able to introduce a corresponding quantum theory which doesn't have the ``chief weakness of quantum theory". This is another motivation of this work. We hope to introduce a quantum theory in which time plays a role like the other three coordinates. The non-relativistic behavior of time is one of the serious obstacles to reconciling quantum mechanics and general relativity. Black holes where both gravitational and quantum effects are important can be used to show how the non-relativistic character of quantum mechanics leads to serious problems. In \cite{Hooft} 't Hooft introduces ``the brick wall model" to reproduce the desired features of a black hole. Although the model is satisfactory, ``it violates the fundamental requirement of coordinate invariance at the horizon". This problem will exist as far as we use a Hamiltonian that is generator of time evolution, and therefore is not invariant under coordinate transformations. His conclusion is even more interesting.`` If the eigenstates of an hamiltonian are used to characterize a Hilbert space then clearly the choice of a coordinate frame affects this characterization. In particular if in a coordinate transformation the quantity $\frac{\partial t'}{\partial t}$ does not everywhere have the same sign the two Hilbert spaces will look very different". We would like to introduce a Hamiltonian which is invariant under coordinate transformations.

In this paper we investigate a five dimensional space without invariance under local Lorentz transformations and examine its consequences. In section II, we introduce a group of transformations under which our work is invariant. Then, transformation laws are given. In section III, we study equations of motion using the five dimensional line element as a lagrangian. We also show that the Lorentz force is involved in the equations of motion. In section IV, we first define connections of this space in a consistent manner. We indicate that a derivative operator defined upon these connections can map tensor fields to tensor fields. Riemann and Ricci tensors are built using the introduced connections. In section V, we write field equations. We, then, show how they separate into the 4-dimensional Einstein and Maxwell equations and that the homogeneous Maxwell equations are included in the theory as $\hat{R}_{(\mu \nu \gamma)4}=0$. In section VI, we introduce a corresponding quantum theory. Spectrum of a free particle and the hydrogen atom are given. In this section we also study Schwarzschild spaces in which a test particle is bound to a massive body due to just gravity. The wave equation of this system is given. The results can be used to determine the thermodynamic properties of a black hole. The conclusions are drawn in section VII.
\section{Frames of Reference}
The first step is to distinguish two concepts. First, coordinate system, and second, frame of reference or observer. We can locate the origin of a coordinate system to any point of our space, but physics only can be described in a group of coordinate systems called frames of reference. Norton best explains the concepts in his paper \cite{Norton} ``In traditional developments of special and general relativity it has been customary not to distinguish between two quite distinct ideas. The first is the notion of a coordinate system, understood simply as the smooth, invertible assignment of four numbers to events in spacetime neighborhoods. The second, the frame of reference, refers to an idealized physical system used to assign such numbers" Anyway, in general relativity we assume general covariance which means all the possible coordinate systems can be carried by observers. As a result, any coordinate system can be a frame of reference. In this regard Bergmann \cite{Bergmann} says ``In the language of four-dimensional geometry, a frame of reference is a four dimensional coordinate system." Unlike other five dimensional theories e.g. Kaluza-Klein, see \cite{overduin} , we don't try to preserve invariance under local Lorentz transformations in our 5D space. We assume, instead, that there exists a 4D invariant plane. Our assumption, therefore, is that physical laws are invariant under the following transformations
\bq
\lb{assumption}
x^{\mu'}=x^{\mu'}(x^{\nu}), ~~~ x^{5'}=x^{5}.
\eq
Here Greek indices run over 0, 1, 2, 3, and $x^5$ indicates the fifth coordinate. This transformation defines a group of coordinate systems within which physics is describable. This group determines all the possible frames of reference or observers. On the other hand, we assume this philosophical doctrine that an observer can only be anything counted as ``matter". It means, matter is observer and observer is matter. We should not discriminate among these two concepts. Thus, by distinguishing all the possible observers, we actually have distinguished all the locations that matter can exist. Equation (\ref{assumption}) not only determines all the possible observers but also determines all the possible locations that matter can exist. Therefore, there can exist no matter in 5D space except on the 4D plane defined by equation (\ref{assumption}). This is the same as saying that our matter can not depend on $x^5$, the fifth coordinate, i.e., partial derivative with respect to the fifth coordinate is zero.

Here we would like to make clear the differences between this theory and Kaluza-Klein (KK) theories. KK theories have invariance under local Lorentz transformations in five dimensions, and in general are invariant under the following transformation
\bq
x^{A'}=x^{A'}(x^{B}).
\eq
Here capital Latin indices run over 0, 1, 2, 3, 5. However, in some KK theories the fifth coordinate is assumed to be compact. In those cases, the fifth coordinate transforms like
\bq
x^{5'}=x^5+f(x^{\mu}).
\eq
These transformations lead to the results mostly different from what we will get later on.
\subsection{Transformation Laws}
Let us first introduce the basis vectors. Like many people who introduce partials $\{\partial_{\mu}\}$ as a basis for the vector space \cite{carroll, hervik} we define partial derivatives with respect to coordinates as basis vectors of the tangent space. The chain rule and transformation in equation (\ref{assumption}) now show us how basis vectors change from an old frame to a new one
\bq
\lb{transformation}
\begin{array}{c}
  \frac{\partial}{\partial x^{\mu'}}=\frac{\partial x^{\mu}}{\partial x^{\mu'}}\frac{\partial}{\partial x^{\mu}},
 \\
 * \\
  \frac{\partial}{\partial x^{5'}}=\frac{\partial}{\partial x^{5}}.
\end{array}
\eq
Vectors must remain unchanged in a change of frame, so we can derive transformation law for vector components. Using equation (\ref{transformation}) we have
\bq
\hat{v}^{A}e_{A}=\hat{v}^{A'}e_{A'}=\hat{v}^{\mu'}e_{\mu'}+\hat{v}^{5'}e_{5'}=\hat{v}^{\mu'}\frac{\partial x^{\mu}}{\partial x^{\mu'}}e_{\mu}+\hat{v}^{5'}e_{5},
\eq
or
\bq
\begin{array}{c}
\lb{ctransformation}
  \hat{v}^{\mu}=\hat{v}^{\mu'}\frac{\partial x^{\mu}}{\partial x^{\mu'}}, \\
  * \\
   \hat{v}^{5}=\hat{v}^{5'}.
\end{array}
\eq
Here ``~$\hat{}$~" indicates a five dimensional object while four dimensional quantities have no ``~$\hat{}$~".

Just like partial derivatives of coordinates which make basis vectors of the tangent space, gradients of coordinates can also form basis one-forms for the cotangent space. Again, the chain rule and transformations in (\ref{assumption}) show how basis one-forms transform under change of frames
\bq
\begin{array}{c}
  dx^{\mu}=dx^{\mu'} \frac{\partial x^{\mu}}{\partial x^{\mu'}}, \\
   \\
  dx^{5}=dx^{5'}.
\end{array}
\eq
Then, form components transform as follows
\bq
\lb{contransformation}
\begin{array}{c}
\hat{\omega}_{\mu}=\hat{\omega}_{\mu'} \frac{\partial x^{\mu'}}{\partial x^{\mu}}, \\
\\
\hat{\omega}_{5}=\hat{\omega}_{5'}.
\end{array}
\eq
The metric tensor is defined as the inner product of basis vectors
\bq
\lb{metric}
\hat{g}_{_{AB}}=\hat{g}(e_{_{A}},e_{_{B}})= e_{_{A}}. e_{_{B}}~.
\eq
Equation (\ref{metric}) determines how metric components transform in a change of frame
\bq
\begin{array}{c}
  \hat{g}_{\mu \nu}=\frac{\partial x^{\mu'}}{\partial x^{\mu}} \frac{\partial x^{\nu'}}{\partial x^{\nu}} \hat{g}_{\mu' \nu'},\\
   \\
  \hat{g}_{\mu 5}=\hat{g}_{\mu' 5'} \frac{\partial x^{\mu'}}{\partial x^{\mu}}, \\
   \\
  \hat{g}_{5 5}=\hat{g}_{5' 5'}.
\end{array}
\eq
\section{Equation of Motion}
Let us consider the following lagrangian
\bq
\lb{lagrangian}
L=\frac{1}{2}m\hat{g}_{AB}\dot{x}^A \dot{x}^B,
\eq
where m is mass of particle, and $\dot{x}^A=\frac{dx^{^A}}{d\lambda}$.
The action, then, is given by
\bq
S=\frac{1}{2}m\int \hat{g}_{AB}\dot{x}^A \dot{x}^B d\lambda.
\eq
Variation of this action gives
\bqn
\lb{variation}
\delta S=\int(\hat{g}_{AB,C}\delta x ^{C}\dot{x}^A \dot{x}^B+2\hat{g}_{AB}\dot{x}^{A}\delta \dot{x}^{B})d\lambda \nb \\
=\int(\hat{g}_{AB,C}\dot{x}^A \dot{x}^B-\frac{d}{d\lambda}(2\hat{g}_{AC}\dot{x}^{A}))\delta x ^{C}d\lambda \nb \\
=0.
\eqn
A consequence of equation (\ref{assumption}) and the statements below it is that $\delta x^{\mu}$ can be chosen arbitrarily, but only $\delta x^{5}=0$ is acceptable. Therefore, equation (\ref{variation}) leads to
\bq
\int(\hat{g}_{AB,\gamma}\dot{x}^A \dot{x}^B-\frac{d}{d\lambda}(2\hat{g}_{A \gamma}\dot{x}^{A}))\delta x ^{\gamma}d\lambda \nb \\
=0,
\eq
or
\bq
\lb{lagrange}
\hat{g}_{AB,\gamma}\dot{x}^A \dot{x}^B-\frac{d}{d\lambda}(2\hat{g}_{A \gamma}\dot{x}^{A})=0,
\eq
which can equivalently be written as
\bq
\hat{g}_{A \gamma}\ddot{x}^{A}+\frac{1}{2}(\hat{g}_{A \gamma, B}+\hat{g}_{B \gamma, A}-\hat{g}_{A B, \gamma})\dot{x}^{A}\dot{x}^{B}=0.
\eq
This equation can be written as follows
\bq
\lb{motionequation}
\ddot{x}^{\alpha}+\frac{1}{2}g^{\alpha \gamma}(\hat{g}_{A \gamma, B}+\hat{g}_{B \gamma, A}-\hat{g}_{A B, \gamma})\dot{x}^{A}\dot{x}^{B}=-g^{\alpha \gamma}\hat{g}_{5 \gamma}\ddot{x}^5,
\eq
where $g^{\alpha \gamma}$ is inverse of four dimensional metric tensor, $g^{\alpha \gamma}\hat{g}_{\gamma \beta}=\delta^{\alpha}_{\beta}$.

Now let's consider the velocity vector, $\dot{x}^A=\frac{dx^{^{A}}}{d\lambda}$. This quantity must obey the transformation law equation (\ref{ctransformation}). We can take $d\lambda=-dx^5$ because $dx^{5'}=dx^5$, then, the velocity vector will be given by
\bq
\lb{velocity}
\dot{x}^{A}=\frac{dx^{A}}{-dx^{5}},
\eq
which transforms as
\bqn
\frac{dx^{\mu'}}{-dx^{5'}}=\frac{\partial x^{\mu'}}{\partial x^{\mu}}\frac{dx^{\mu}}{-dx^{5}}, \nb \\
\frac{dx^{5'}}{-dx^{5'}}=\frac{dx^{5}}{-dx^{5}}=-1.
\eqn
If we substitute equation (\ref{velocity}) into equation (\ref{motionequation}), we'll get the equation of motion
\bq
\lb{motion}
\ddot{x}^{\alpha}+\frac{1}{2}g^{\alpha \gamma}(\hat{g}_{A \gamma, B}+\hat{g}_{B \gamma, A}-\hat{g}_{A B, \gamma})\dot{x}^{A}\dot{x}^{B}=0,
\eq
because, in agreement with Eq. (\ref{velocity}), $\ddot{x}^{5}$ is zero.

Consequently, we are able to define the connections as
\bqn
\lb{connection}
\hat{\Gamma}^{\alpha}_{AB}=\frac{1}{2}g^{\alpha \gamma}(\hat{g}_{A \gamma, B}+\hat{g}_{B \gamma, A}-\hat{g}_{A B, \gamma}), \nb \\
\hat{\Gamma}^{5}_{AB}=0,~~~~~~~~~~~~~~~~~~~~~~~~~~~~~~~~~~~~~
\eqn
provided they satisfy all the required conditions to construct the covariant derivative. In the next section we consider those conditions, and show that they are all satisfied.

The Lorentz force is a natural constituent of the equation of motion. Equation (\ref{motion}) can be written as
\begin{equation}
\hat{\ddot{x}}^{\kappa}+\hat{\Gamma}^{\kappa}_{\alpha \beta}\hat{\dot{x}}^{\alpha}\hat{\dot{x}}^{\beta}-2\hat{\Gamma}^{\kappa}_{\alpha 5}\hat{\dot{x}}^{\alpha}+\hat{\Gamma}^{\kappa }_{5 5}=0.
\end{equation}
Here, in line with equation (\ref{velocity}),  we have used $\hat{\dot{x}}^5=-1$.
The equation of motion then is given by
\bqn
\lb{eq19}
\hat{\ddot{x}}^{\kappa}+\frac{1}{2}g^{\lambda
\kappa}(\hat{g}_{\lambda\beta,\alpha}+\hat{g}_{\lambda\alpha,\beta}-
\hat{g}_{\alpha\beta,\lambda})\hat{\dot{x}}^{\alpha}\hat{\dot{x}}^{\beta}\nb\\
-g^{\lambda\kappa}(\hat{g}_{5 \lambda,\alpha}-\hat{g}_{5\alpha,\lambda})\hat{\dot{x}}^{\alpha}-\frac{1}{2}g^{\lambda \kappa}\hat{g}_{55,\lambda}=0,
\eqn
knowing that partial derivatives with respect to $\hat{x}^{5}$ are zero, and a
comma  indicates partial derivative. At this moment we can define the vector potential,
$A_{\mu}$, as
\begin{equation}
\lb{metric5}
A_{\mu} \equiv \frac{m}{q}\hat{g}_{5 \mu},
\end{equation}
where $m$ and $q$ are the mass and electric charge of our test
particle respectively. Therefore, we define the Faraday tensor as
\begin{equation}
\lb{faraday}
F_{\mu \nu} \equiv \hat{g}_{5 \mu,\nu}-\hat{g}_{5 \nu,\mu}.
\end{equation}
Then, the second and third terms in equation (\ref{eq19}), can be interpreted as the gravitational and Lorentz forces respectively.

\section{Curvature}
Since the partial derivative of a tensor doesn't transform like a tensor, we need to build a covariant derivative. This derivative operator is required to have some properties which are expressed for example in \cite{wald}. To meet these requirements, we should write the covariant derivative as
\bq
\lb{dkaluza}
(\hat{\nabla}_{A}\hat{f})^{B}=\partial_{A}\hat{f}^{B}+\hat{\Gamma}^{B}_{AC}\hat{f}^{C}.
\eq
Now we should define connections, $\hat{\Gamma}^{B}_{AC}$, so that this derivative operator transforms like a tensor.
Under transformation (\ref{assumption}), an arbitrary tensor, say $\hat{T}^{A}_{B}$, must follow the transformation laws (\ref{ctransformation}) and (\ref{contransformation}) which can be written as
\bqn
\lb{tlaw2}
\hat{T}^{\alpha}_{\beta}=\frac{\partial x^{\alpha}}{\partial x^{\alpha'}}\frac{\partial x^{\beta'}}{\partial x^{\beta}}\hat{T}^{\alpha'}_{\beta'},\nb\\
\hat{T}^{\alpha}_{5}=\frac{\partial x^{\alpha}}{\partial x^{\alpha'}}\hat{T}^{\alpha'}_{5'},~~~~\nb\\
\hat{T}^{5}_{\beta}=\frac{\partial x^{\beta'}}{\partial x^{\beta}}\hat{T}^{5'}_{\beta'},~~~~\nb\\
\hat{T}^{5}_{5}=\hat{T}^{5'}_{5'}.~~~~~~~~
\eqn
Kaluza-Klein five dimensional connections which are defined as
\bq
\lb{ckaluza}
\hat{\Gamma}^{C}_{AB}=\frac{1}{2}\hat{g}^{C D}(\hat{g}_{A D, B}+\hat{g}_{B D, A}-\hat{g}_{A B, D}),
\eq
meet these requirements. But, as we all know they can't give rise to the two homogeneous Maxwell equations. We also know that they are not unique connections. Wald \cite{wald} in this regard says ``given only the manifold structure, many distinct derivative operators are possible and no one of them is naturally preferred over the others".
If we instead utilize connections introduced in (\ref{connection}) to construct (\ref{dkaluza}), all the Maxwell equations can be reached. It also will transform in agreement with the transformation laws (\ref{tlaw2})
\bqn
(\hat{\nabla}_{\alpha}\hat{f})^{\beta}=\frac{\partial x^{\alpha'}}{\partial x^{\alpha}}\frac{\partial x^{\beta}}{\partial x^{\beta'}}(\hat{\nabla}_{\alpha'}\hat{f})^{\beta'}, \nb \\
* \nb \\
(\hat{\nabla}_{5}\hat{f})^{\beta}=\frac{\partial x^{\beta}}{\partial x^{\beta'}}(\hat{\nabla}_{5'}\hat{f})^{\beta'},~~~~ \nb \\
* \nb \\
(\hat{\nabla}_{\beta}\hat{f})^{5}=\frac{\partial x^{\beta'}}{\partial x^{\beta}}(\hat{\nabla}_{\beta'}\hat{f})^{5'},~~~~ \nb \\
* \nb \\
(\hat{\nabla}_{5}\hat{f})^{5}=(\hat{\nabla}_{5'}\hat{f})^{5'}.~~~~~~~~
\eqn
One more thing about the connections (\ref{connection}) is that replacing $A$ and $B$ by $\alpha$ and $\beta$ will lead to connections belonging to general relativity
\bqn
\hat{\Gamma}^{\lambda}_{\alpha \beta}=\frac{1}{2}g^{\lambda \gamma}(\hat{g}_{\alpha \gamma, \beta}+\hat{g}_{\beta \gamma, \alpha}-\hat{g}_{\alpha \beta, \gamma}) \nb \\
=\frac{1}{2}g^{\lambda \gamma}(g_{\alpha \gamma, \beta}+g_{\beta \gamma, \alpha}-g_{\alpha \beta,\gamma}),
\eqn
which equals $\Gamma^{\lambda}_{\alpha \beta}$.\\
The $5D$ curvature tensor, $\hat{R}^{^D}_{_{ABC}}$, is defined by
\begin{equation}
\hat{\nabla}_{A}\hat{\nabla}_{B}\hat{f}_{C}-\hat{\nabla}_{B}\hat{\nabla}_{A}\hat{f}_{C}=\hat{R}^{^D}_{_{ABC}}\hat{f}_{D},\label{ab}
\end{equation}
where $\hat{f}_{C}$ is an arbitrary $5D$ vector. Therefore, the Riemann curvature tensor will be
\begin{equation}
\hat{R}^{^{D}}_{_{CBA}}=\partial_{B}\hat{\Gamma}^{D}_{AC}-\partial_{A}\hat{\Gamma}^{D}_{BC}+\hat{\Gamma}^{D}_{B
E}\hat{\Gamma}^{E}_{AC}-\hat{\Gamma}^{D}_{A E}\hat{\Gamma}^{E}_{B
C}\label{a5}.
\end{equation}
This equation will separate into
\begin{equation}
\label{riemann}
\hat{R}^{\kappa}_{_{CBA}}=\partial_{B}\hat{\Gamma}^{\kappa}_{AC}-\partial_{A}\hat{\Gamma}^{\kappa}_{BC}+\hat{\Gamma}^{\kappa}_{B
\gamma}\hat{\Gamma}^{\gamma}_{AC}-\hat{\Gamma}^{\kappa}_{A
\gamma}\hat{\Gamma}^{\gamma}_{B C}\label{a6},
\end{equation}
and
\begin{equation}
\label{riemann5}
\hat{R}^{5}_{_{CBA}}=0\label{a7}.
\end{equation}
Here again it is clear that $\hat{R}^{\kappa}_{\alpha \beta
\gamma}=R^{\kappa}_{\alpha \beta \gamma}$ where $R^{\kappa}_{\alpha \beta
\gamma}$ is the $4D$ Riemann tensor. Now let's show that
\begin{equation}
\hat{R}^{\alpha}_{\beta \kappa
5}=\frac{1}{2}\nabla_{\kappa}F^{\alpha}_{\beta},\label{eq33}
\end{equation}
where $F_{\alpha \beta}$ is the Faraday tensor given by equation (\ref{faraday}). We write
\begin{equation}
\frac{1}{2}F^{\kappa}_{\sigma}=\frac{1}{2}g^{\kappa
\lambda}(\hat{g}_{5 \lambda,\sigma}-\hat{g}_{5
\sigma,\lambda})=\hat{\Gamma}^{\kappa}_{5\sigma},
\end{equation}
so
\begin{equation}
\frac{1}{2}\nabla_{\kappa}F^{\alpha}_{\beta}=\partial_{\kappa}\hat{\Gamma}^{\alpha}_{5\beta}+
\Gamma^{\alpha}_{\kappa\lambda}\hat{\Gamma}^{\lambda}_{5\beta}-\Gamma^{\lambda}_{\kappa
\beta}\hat{\Gamma}^{\alpha}_{5\lambda}.
\end{equation}
On the other hand
\bqn
\hat{R}^{\alpha}_{\beta \kappa
5}=\partial_{\kappa}\hat{\Gamma}^{\alpha}_{5\beta}+\hat{\Gamma}^{\alpha}_{\kappa
\lambda}\hat{\Gamma}^{\lambda}_{5\beta}-\hat{\Gamma}^{\lambda}_{\kappa
\beta}\hat{\Gamma}^{\alpha}_{5\lambda}, \nb \\
=\partial_{\kappa}\hat{\Gamma}^{\alpha}_{5\beta}+
\Gamma^{\alpha}_{\kappa\lambda}\hat{\Gamma}^{\lambda}_{5\beta}-\Gamma^{\lambda}_{\kappa
\beta}\hat{\Gamma}^{\alpha}_{5\lambda},
\eqn
hence
\begin{equation}
\hat{R}^{\alpha}_{\beta \kappa
5}=\frac{1}{2}\nabla_{\kappa}F^{\alpha}_{\beta}\label{a8}.
\end{equation}
The Ricci tensor is
\begin{equation}
\hat{R}_{_{CA}}=\hat{R}^{^{B}}_{_{CBA}}\label{a5},
\end{equation}
or
\begin{equation}
\lb{ricci}
\hat{R}_{_{CA}}=\partial_{\kappa}\hat{\Gamma}^{\kappa}_{AC}-\partial_{A}\hat{\Gamma}^{\kappa}_{\kappa
C}+\hat{\Gamma}^{\kappa}_{\kappa
\gamma}\hat{\Gamma}^{\gamma}_{AC}-\hat{\Gamma}^{\kappa}_{A
\gamma}\hat{\Gamma}^{\gamma}_{\kappa C}.
\end{equation}
Again
\begin{equation}
\hat{R}_{\alpha \gamma}=R_{\alpha \gamma},
\end{equation}
where $R_{\alpha \gamma}$ is the $4D$ Ricci tensor. It is easy to show that $\hat{R}_{AB}$ is symmetric. We can also see from equation
(\ref{ricci})  that
\begin{equation}
\lb{ricci5}
\hat{R}_{5 \lambda}=\frac{1}{2}\nabla_{\alpha}F^{\alpha}_{\lambda}.
\end{equation}
The remaining component of the Ricci tensor is
\bq
\hat{R}_{55}=-\frac{1}{2}\nabla_{\mu}\nabla^{\mu}\hat{g}_{55}-\frac{1}{4}F^{\mu \nu}F_{\nu \mu}.
\eq

The 5D Ricci scalar is $\hat{R}=\hat{g}^{CA} \hat{R}_{CA}$. But, another scalar also can be defined as $R=g^{\alpha \beta} \hat{R}_{\alpha \beta}$. They both
transform like a scalar under the transformation (\ref{assumption}).
\section{Field Equations}
To derive the 4D Einstein and Maxwell equations we must write the 5D field equations as
\begin{equation}
\lb{fieldequation}
\hat{R}_{_{AB}}-\frac{1}{2}R \hat{g}_{_{AB}}=\kappa
\hat{T}_{_{AB}},\label{ac}
\end{equation}
where $R$ is defined as $R=g^{\alpha \beta} \hat{R}_{\alpha \beta}$, and $\hat{T}_{AB}$ is
\begin{equation}
\lb{TAB}
\hat{T}_{AB}=\left(%
\begin{array}{ccccc}
  T_{\alpha\beta} &   &   &   & \frac{q}{m}\epsilon j_{0} \\
    &   &   &   & \frac{q}{m}\epsilon j_{1} \\
    &   &   &   & \frac{q}{m}\epsilon j_{2} \\
    &   &   &   & \frac{q}{m}\epsilon j_{3} \\
  \frac{q}{m}\epsilon j_{0} & \frac{q}{m}\epsilon j_{1} & \frac{q}{m}\epsilon j_{2} & \frac{q}{m}\epsilon j_{3} & \hat{T_{55}} \\
\end{array}%
\right).
\end{equation}
Here $\epsilon$ is a coupling constant, $T_{\alpha \beta}$ is the 4D energy momentum tensor, and $j_{\mu}$ is the 4D electromagnetic current vector.
Now if we replace $A$ by $\alpha$ and $B$ by $\beta$, equation (\ref{fieldequation}) yields
\begin{equation}
\hat{R}_{_{\alpha \beta}}-\frac{1}{2} R \hat{g}_{_{\alpha \beta}}=\kappa
\hat{T}_{_{\alpha \beta}},
\end{equation}
or equivalently
\begin{equation}
R_{_{\alpha \beta}}-\frac{1}{2} R g_{_{\alpha \beta}}=\kappa
T_{_{\alpha \beta}}.
\end{equation}
The left hand side of this equation is the $4D$ Einstein tensor, and we all know that $\nabla^{\alpha}(R_{_{\alpha \beta}}-\frac{1}{2} R g_{_{\alpha \beta}})=0$. As a result,
\bq
\nabla^{\alpha}T_{\alpha \beta}=0,
\eq
which represents conservation of energy-momentum. Alternatively, if we replace $A$ by $5$ and $B$ by $\beta$, equation (\ref{fieldequation}) leads to the following equation
\begin{equation} \hat{R}_{5 \beta}-\frac{1}{2}\hat{g}_{5 \beta}R=\kappa
\hat{T}_{5 \beta},
\end{equation}
which using equations (\ref{ricci5}), and (\ref{TAB}) results in Maxwell equations with a correction term
\begin{equation}
\nabla_{\alpha}F^{\alpha}_{\beta}-\hat{g}_{5 \beta}R=\alpha j_{\beta},
\end{equation}
where $\alpha$ is a constant. We know that $\nabla^{\beta}\nabla_{\alpha}F^{\alpha}_{\beta}=0$, so what is conservative here is
\bq
\lb{current}
 {\cal{J}}^{\beta}=j^{\beta}+c A^{\beta} R,
\eq
where $c$ is a constant, and $A^{\beta}$ comes from equation (\ref{metric5}).
We call it the electromagnetic current vector in the presence of gravity. And,
\bq
\nabla_{\beta}{\cal{J}}^{\beta}=0.
\eq
It is not unusual that the conservative current vector changes when a second field comes to exist with the initial field. Let's give an example that has nothing to do with this work, but is still a good analog.

Consider a complex Klein-Gordon field in a flat space-time in the absence of electromagnetism. The conservative charge-current vector is, see for example \cite{Mandl},
\bq
j_{_{_{KG}}}^{\beta}=-iq(\phi \frac{\partial \phi^{\dag}}{\partial x_{\beta}}-\phi^{\dag}\frac{\partial \phi}{\partial x_{\beta}} ).
\eq
Here $q$ is charge. In this case we have
\bq
\partial_{\beta}j_{_{_{KG}}}^{\beta}=0.
\eq
But, when both Klein-Gordon and electromagnetic fields exist, the conservative charge-current vector is, see for \\
example \cite{Guidry},
\bq
\lb{kleingordoncurrent}
{\cal{J}}_{_{_{KG}}}^{\beta}=j_{_{_{KG}}}^{\beta}-2q^2 A^{\beta} \phi \phi^{\dag},
\eq
where
\bq
\partial_{\beta}{\cal{J}}_{_{_{KG}}}^{\beta}=0.
\eq
The interesting point is that in both equations (\ref{current}) and (\ref{kleingordoncurrent}) the extra term is proportinal to the vector potential, $A^{\beta}$.

So far we have derived the non-homogeneous Maxwell equations. As is well known, the homogeneous Maxwell
equations appear as constraints on the Faraday tensor in the KK theory and are
implicitly assumed to hold. They do do not appear in an independent manner.
However, the situation is different in the theory presented here. It is therefore
appropriate at this point to show that this is indeed the case. To begin with we
note that in classical electrodynamics, the Jacobi identities lead to the
homogeneous Maxwell equations if we define our connections as
$D_{\mu}=\partial_{\mu}+A_{\mu}$, where $A_{\mu}$ is the four vector potential,
so that the Faraday tensor is written as $F_{\mu \nu}=[D_{\mu},D_{\nu}]$. Now,
use of the second Jacobi identity leads to the Bianchi identity in the form
\begin{equation}
[D_{\mu},F_{\nu \lambda}]+[D_{\nu},F_{\lambda
\mu}]+[D_{\lambda},F_{\mu \nu}]=0,
\end{equation}
which is equivalent to
\begin{equation}
D_{\mu}F_{\nu \lambda}+D_{\nu}F_{\lambda \mu}+D_{\lambda}F_{\mu
\nu}=0.
\end{equation}
This is, of course,  the homogeneous Maxwell equations. For details, we refer readers to \cite{Baez}. In General
Relativity on the other hand, we replace $D_{\mu}$ by $\nabla_{\mu}$, so that
the first Jacobi identity results in
\begin{equation}
[\nabla_{\mu},R(\partial_{\nu},
\partial_{\lambda})]+[\nabla_{\nu},R(\partial_{\lambda},
\partial_{\mu})]+[\nabla_{\lambda},R(\partial_{\mu},
\partial_{\nu})]=0,
\end{equation}
or
\begin{equation}
R_{\alpha \beta [\mu \nu ;\lambda]}=0.\label{eq100}
\end{equation}
The torsion free condition also provides the first Bianchi identity
\begin{equation}
R_{\lambda [\beta \gamma \delta]}=0.\label{eq101}
\end{equation}
It is now clear that because of the definition of connections in general relativity,
neither the first Bianchi identity, equation (\ref{eq101}), nor the second Bianchi
identity, equation (\ref{eq100}), lead to the homogeneous Maxwell equations. Let us
now show that in the model presented here, the first Bianchi identity leads to the
homogeneous Maxwell equation. The first Bianchi identity can be written as
\begin{equation}
\label{identity}
\hat{R}_{\alpha \beta \gamma 5}+\hat{R}_{\gamma \alpha \beta
5}+\hat{R}_{\beta \gamma \alpha 5}=0,
\end{equation}
where $\hat{R}_{_{ABCD}}=\hat{g}_{_{AE}}\hat{R}^{^E}_{_{BCD}}$. One can easily prove this identity by substituting equations (\ref{riemann}) and (\ref{riemann5}) in equation (\ref{identity}). Now using equation (\ref{a8}), the first Bianchi identity, equation (\ref{identity}), leads to
\begin{eqnarray}
\nabla_{\kappa}F_{\alpha
\beta}+\nabla_{\beta}F_{\kappa
\alpha}+\nabla_{\alpha}F_{\beta \kappa}=0, \label{eq39}
\end{eqnarray}
showing that, as expected, the first Bianchi identity results in the homogeneous
Maxwell equations. It is worth mentioning that in other $5D$ theories, e.g. KK,
Bianchi identities do not lead to equation (\ref{eq39}).
\\

The remaining component of equation (\ref{fieldequation}) is
\bq
\lb{g55}
-\frac{1}{2}\nabla_{\mu}\nabla^{\mu}\hat{g}_{55}-\frac{1}{4}F^{\mu \nu}F_{\nu \mu}-\frac{1}{2}R\hat{g}_{55}=\kappa \hat{T}_{55}.
\eq
In the case of an empty space this leads to
\bq
\nabla_{\mu}\nabla^{\mu}\hat{g}_{55}=\frac{1}{2}F^{\mu \nu}F_{\mu \nu}.
\eq
This is to some extent similar to the corresponding Kaluza-Klein equation
\bq
\nabla_{\mu}\nabla^{\mu}\hat{g}_{55}=\frac{\kappa^2 \hat{g}_{55}^3}{4}F^{\mu \nu}F_{\mu \nu},
\eq
with this difference that in our work $\hat{g}_{55}$ does not need to be a constant in order to get the Einstein equations.
\section{Quantum Mechanics}
In the preceding sections we have presented a classical theory. At this time, we should use ``classical analogy" to construct a corresponding quantum theory. In section III we considered motion of particles. As it is apparent in that section, the dynamical variables are the canonical coordinates, $x^{\mu}$ and momenta, $p_{\mu}$. It is obvious that $x^{5}$ can not be a dynamical variable because as shown above $\delta x^5$ equals zero, and equation of motion (\ref{motion}) can be derived from the Lagrange equation in (\ref{lagrangeequation}) by taking only $x^{\mu}$ and its momenta as dynamical variables. Taking (\ref{lagrangian}) as lagrangian, equation (\ref{lagrange}) is equal to the following one
\bq
\lb{lagrangeequation}
\frac{d}{-dx^5}(\frac{\partial L}{\partial \dot{x}^{\mu}})-\frac{\partial L}{\partial x^{\mu}}=0,
\eq
where $\dot{x}^{\mu}=\frac{dx^{\mu}}{-dx^5}$. In this section we are more interested in the Hamiltonian equation of motion since we want to get in the quantum realm.

The conjugate momenta are defined as
\bq
\lb{momentum}
p_{\mu} \equiv \frac{\partial L}{\partial \dot{x}^{\mu}},
\eq
where $\dot{x}^{\mu}=\frac{dx^{\mu}}{-dx^5}$. Then, from equation (\ref{lagrangeequation}) we have
\bq
\dot{p}_{\mu} \equiv \frac{d p_{\mu}}{-dx^5}=\frac{\partial L}{\partial x^{\mu}}.
\eq
A Legendre transformation generates our Hamiltonian
\bq
\lb{HD}
H=\dot{x}^{\mu}p_{\mu}-L.
\eq
The equations of motion, then, are
\bqn
\lb{hamiltoneq}
\frac{\partial H}{\partial x^{\mu}}=-\dot{p}_{\mu},\nb\\
\frac{\partial H}{\partial p_{\mu}}=\dot{x}^{\mu},~~
\eqn
where a dot indicates $\frac{d}{-dx^5}$. Differentiation of an arbitrary function, say $Q(x^{\mu},p_{\mu})$, with respect to $-x^5$ is
\bq
\lb{differential}
\frac{d Q}{-dx^5}=\frac{\partial Q}{\partial x^{\mu}}\dot{x}^{\mu}+\frac{\partial Q}{\partial p_{\mu}}\dot{p}_{\mu}.
\eq
Substituting $\dot{x}^{\mu}$ and $\dot{p}_{\mu}$ from equation (\ref{hamiltoneq}) into this equation gives
\bq
\frac{d Q}{-dx^5}=\frac{\partial Q}{\partial x^{\mu}}\frac{\partial H}{\partial p_{\mu}}-\frac{\partial Q}{\partial p_{\mu}}\frac{\partial H}{\partial x^{\mu}},
\eq
which is the Poisson Bracket of $Q$ and $H$
\bq
\lb{PBH}
\Big\{Q,H\Big\}=\frac{d Q}{-dx^5}.
\eq
To derive the quantum rules, we may refer to Dirac's famous book \cite{Dirac}. Quantum mechanical relations can be obtained from the corresponding classical relations just by replacing classical Poisson brackets by commutators as follows
\bq
\Big\{~~,~\Big\}\longrightarrow \frac{[~~,~]}{i\hbar}.
\eq
Then, for example, from equation (\ref{PBH}) we can write
\bq
\lb{QPBH}
\Big[\textbf{Q},\textbf{H}\Big]=i\hbar\frac{d \textbf{Q}}{-dx^5},
\eq
which obviously is written in the Heisenberg picture. Here bold letters indicate operators.

\subsection{Wave Equation}
To deduce the wave equation, it would be more convenient to expand our lagrangian, equation (\ref{lagrangian}),
\bq
L=\frac{1}{2}m\hat{g}_{\mu \nu}\dot{x}^{\mu}\dot{x}^{\nu}-m\hat{g}_{5 \mu}\dot{x}^{\mu}+\frac{1}{2}m\hat{g}_{55}.
\eq
By use of equation ({\ref{metric5}}) we would like to replace $\hat{g}_{5 \mu}$ by $\frac{q}{m}A_{\mu}$. Thus,
\bq
L=\frac{1}{2}m\hat{g}_{\mu \nu}\dot{x}^{\mu}\dot{x}^{\nu}-q A_{\mu}\dot{x}^{\mu}+\frac{1}{2}m\hat{g}_{55}.
\eq
The conjugate momenta, equation (\ref{momentum}), are
\bq
p_{\mu}=m\hat{g}_{\mu \nu}\dot{x}^{\nu}-q A_{\mu}.
\eq
The Hamiltonian, equation (\ref{HD}), then, becomes
\bq
\lb{HQU}
H=\frac{1}{2m}(p_{\mu}+q A_{\mu})(p^{\mu}+q A^{\mu})-\frac{1}{2}m\hat{g}_{55}.
\eq
Keeping equation (\ref{QPBH}) in mind the wave equation is
\bq
\lb{waveeq}
\textbf{H}\psi=i\hbar\frac{\partial \psi}{-\partial x^5},
\eq
provided momentums, $\textbf{p}_{\mu}$, are replaced by the following differential operators
\bq
\textbf{p}_{\mu}\longrightarrow -i\hbar\nabla_{\mu}.
\eq
Using (\ref{HQU}, \ref{waveeq}), the wave equation takes the form
\bqn
\lb{waveequation}
\frac{-\hbar^2}{2m}\nabla_{\mu}\nabla^{\mu} \psi-\frac{i\hbar q}{2m}(\nabla_{\mu}A^{\mu})\psi-\frac{i\hbar q}{m}A^{\mu}\nabla_{\mu}\psi\nb\\
+\frac{q^2}{2m}A_{\mu}A^{\mu}\psi-\frac{1}{2}m\hat{g}_{55}\psi=i\hbar\frac{\partial \psi}{-\partial x^5}.~~~~~~~~~
\eqn
\subsection{Free Particle}
To deal with free particles, we should assume a flat space where neither gravity nor electromagnetism exist. Therefore, $g_{\mu \nu}=\eta_{\mu \nu}$, and $A^{\mu}=0$. Using equation (\ref{g55}) we can conclude that $\hat{g}_{55}$ is a constant in this case. We assume this constant is so small that can be neglected. The wave equation (\ref{waveequation}), thus, becomes
\bq
\frac{-\hbar^2}{2m}\partial^\mu\partial_\mu\psi=i\hbar\frac{\partial \psi}{-\partial x^5},
\eq
or
\bq
\frac{-\hbar^2}{2m}(\frac{\partial^2 \psi}{c^2 \partial t^2}-\nabla^2 \psi)=i\hbar\frac{\partial \psi}{-\partial x^5}.
\eq
This equation has a solution of the form
\bq
\psi=exp~i(\vec{k}.\vec{r}-\omega t)exp(\frac{iZx^5}{\hbar}).
\eq
when the following relation is satisfied
\bq
\hbar^2\omega^2=\hbar^2 c^2 k^2+2mc^2Z.
\eq
Comparing with the well known equation
\bq
\hbar^2\omega^2=\hbar^2 c^2 k^2+m^2c^4,
\eq
one can conclude that $Z$ is half the rest energy of the particle
\bq
\lb{H}
Z=\frac{1}{2}mc^2.
\eq
\subsection{Hydrogen Atom}
In this part we would like to derive the Hydrogen spectrum. For this purpose we again assume a flat space where there isn't any gravitational field. We also replace $q$ by $-e$ where $e$ is the charge of an electron. The wave equation (\ref{waveequation}) in this case reads
\bqn
\lb{Hg55}
\frac{-\hbar^2}{2m}\partial_{\mu}\partial^{\mu} \psi+\frac{i\hbar e}{2m}(\partial_{\mu}A^{\mu})\psi+\frac{i\hbar e}{m}A^{\mu}\partial_{\mu}\psi\nb\\
+\frac{e^2}{2m}A_{\mu}A^{\mu}\psi-\frac{1}{2}m\hat{g}_{55}\psi=i\hbar\frac{\partial \psi}{-\partial x^5}.~~~
\eqn
To deal with $\hat{g}_{55}$ term we must solve equation (\ref{g55}). Here the only non-zero component of the Faraday tensor is $F_{tr}$ which equals the radial electric field. We assume a spherically symmetric case and solve the resulting Laplace equation. The solution shows that the $\hat{g}_{55}$ term in equation (\ref{Hg55}) is smaller than the other terms, by at least $-20$ orders of magnitude. So, it can be dropped.
A solution of the following form can satisfy equation (\ref{Hg55})
\bq
\psi=u(r, \theta, \phi)exp(-i\omega t)exp(\frac{i Z x^5}{\hbar}).
\eq
Here $Z$ is the same as that of a free particle
\bq
Z=\frac{1}{2}mc^2.
\eq
If we also substitute $A^{\mu}=(\frac{\phi(r)}{c},\vec{0})$ into the wave equation, it gives
\bqn
\lb{hydrogen}
(-\hbar^2 c^2 \nabla^2+m^2c^4)u(r, \theta, \phi)=(\hbar \omega+e\phi(r))^2u(r, \theta, \phi),\nb\\
*\nb\\
e>0.~~~~~~~~~~~~~~~~~~~~~~
\eqn
This is Schrodinger's relativistic wave equation for the Hydrogen atom, and leads to the following spectrum for the hydrogen atom
\bq
\omega=\frac{mc^2}{\hbar}\Big[1-\frac{\gamma^2}{2n^2}-\frac{\gamma^4}{2n^4}(\frac{n}{l+\frac{1}{2}}-\frac{3}{4})\Big],
\eq
if we take a Coulomb field. We have used Schiff's book \cite{Schiff} to write this spectrum. One can also compare equation (\ref{hydrogen}) with the related equation in this book.
\subsection{Schwarzschild Spaces}
In this section we are interested in the quantum structure of the systems described by the Schwarzschild metric. Our assumption is that two particles bound together due to gravitational force should have some quantum mechanical states. The situation here is similar to that of the hydrogen atom. But, the interaction forces are different. The Schwarzschild metric is
\bq
\hat{g}_{\mu \nu}=\left(
                    \begin{array}{cccc}
                      -(1-\frac{2M}{r}) &  &  &  \\
                       & \frac{1}{(1-\frac{2M}{r})} &  &  \\
                       &  & r^2 &  \\
                       &  &  & r^2 \sin^2\theta \\
                    \end{array}
                  \right).
\eq
We assume $A^{\mu}$ to be zero in this space. Equation (\ref{g55}) also shows that $\hat{g}_{55}$ is a constant which will be added to the Hamiltonian eigenvalue that is another constant. To simplify the situation we just drop the term. Wave equation (\ref{waveequation}), then, becomes
\bq
\lb{HSch}
\frac{-\hbar^2}{2m}\nabla_{\mu}\nabla^{\mu} \psi=i\hbar\frac{\partial \psi}{-\partial x^5}.
\eq
After doing some calculation, this equation leads to
\bqn
-\frac{\hbar^2}{2m}\Big[\frac{-1}{(1-\frac{2M}{r})}\frac{\partial^2}{\partial t^2}+(1-\frac{2M}{r})\frac{\partial^2}{\partial r^2}~~\nb\\
+2(\frac{1}{r}-\frac{M}{r^2})\frac{\partial}{\partial r}+\frac{1}{r^2}\frac{\partial^2}{\partial \theta^2}+\frac{\cos\theta}{r^2 \sin\theta}\frac{\partial}{\partial \theta}\nb\\
+\frac{1}{r^2\sin^2\theta}\frac{\partial^2}{\partial \phi^2}  \Big]\psi=i\hbar\frac{\partial \psi}{-\partial x^5}.~~~~~~~~~~~~
\eqn
This equation has a solution of the following form
\bq
\psi=u(r, \theta, \phi)exp(-i\omega t)exp(\frac{iZx^5}{\hbar}),
\eq
in which $Z$ has the same value as it had in the case of a free particle and Hydrogen atom.
Therefore, this equation yields
\bqn
\lb{schwarzschild}
\nabla^2u(r, \theta, \phi)-\frac{2M}{r}(\frac{\partial^2u(r, \theta, \phi)}{\partial r^2}+\frac{1}{r}\frac{\partial u(r, \theta, \phi)}{\partial r})\nb\\
+(\frac{\omega^2}{1-\frac{2M}{r}}+\frac{m^2c^2}{\hbar^2})u(r, \theta, \phi)=0.~~~~~~
\eqn
$u(r, \theta, \phi)$ also can be separated
\bq
u(r, \theta, \phi)=R(r)Y(\theta,\phi).
\eq
$Y(\theta,\phi)$ in this equation is the same as $Y(\theta,\phi)$ in the hydrogen case. After separation of variables it reads
\bqn
(1-\frac{2M}{r})\frac{\partial^2R}{\partial r^2}+\frac{2}{r}(1-\frac{M}{r})\frac{\partial R}{\partial r}~~~~~\nb\\
+(\frac{\omega^2}{1-\frac{2M}{r}}+\frac{m^2c^2}{\hbar^2}-\frac{l(l+1)}{r^2})R=0,
\eqn
which can be written as following
\bqn
(1-\frac{2M}{r})^{-1}\omega^2R+\frac{1}{r^2}\frac{\partial}{\partial r}(r(r-2M)\frac{\partial R}{\partial r})\nb\\
-(\frac{l(l+1)}{r^2}-\frac{m^2c^2}{\hbar^2})R=0.~~~~~~~~~
\eqn
This is the wave equation 't Hooft uses in \cite{Hooft} to get ``energy" levels of a black hole. To find the thermodynamic properties of the system we need to know the number of ``energy" levels. In \cite{Hooft} using WKB approximation the number of radial modes associated with this equation is given, and then it is used to calculate some thermodynamic quantities of a black hole. Here we stress again that the hamiltonian (\ref{HQU}) and consequently the wave equation (\ref{HSch}) are not observer dependent.
\\
\\
\section{Conclusions}
In this paper we have studied a five dimensional space which is not invariant under local Lorentz transformations. It has been shown that if we, instead, force the theory to remain invariant under a specific group of transformations, we can achieve the four dimensional Einstein and Maxwell equations with the homogeneous Maxwell equations appearing naturally. We also have used the 5-dimensional line element as a lagrangian, and shown that the hamiltonian produced by the Legendre transformation is the generator of $x^5$ evolution. Using this result we have developed a corresponding quantum theory. As examples we have recovered spectrum of a free particle and the hydrogen atom. Moreover, Schwarzschild spaces have been considered. The results can be used to determine some thermodynamic quantities of the system.
\begin{acknowledgments}
We would like to express our thanks to H. R. Sepangi, and A. Wang for useful discussions and valuable suggestions.
\end{acknowledgments}

\end{document}